\documentclass[12pt]{iopart}
\usepackage{graphicx}   

\usepackage{array}

\newcolumntype{C}[1]{>{\centering\arraybackslash$}m{#1}<{$}}
\newlength{\mycolwd}                                         
\newlength{\mycolwdd}
\begin{document}

\title[]{First-principles discovery of stable two-dimensional materials with high-level piezoelectric response}

\author{Tu\u{g}bey Kocaba\c{s}$^1$, Deniz \c{C}ak{\i}r$^2$,
Cem Sevik$^3$}
\address{$^1$ Department of Materials Science and Engineering, Institute of Graduate Programs, Eskisehir Technical University, Eskişehir, TR 26555, Turkey}
\address{$^2$ Department of Physics and Astrophysics, University of North Dakota, Grand Forks, North Dakota 58202, United States}
\address{$^3$ Department of Mechanical Engineering, Faculty of Engineering, Eskisehir Technical University, Eskişehir, TR 26555, Turkey}
\ead{csevik@eskisehir.edu.tr}

\vspace{12pt}
\begin{indented}
\item[]September 2020
\end{indented}

\begin{abstract}
The rational design of two-dimensional piezoelectric materials has recently garnered great interest due to their increasing use in technological applications, including sensor technology, actuating devices, energy harvesting, and medical applications. Several materials possessing high piezoelectric response have been reported so far, but a high-throughput first-principles approach to estimate the piezoelectric potential of layered materials has not been performed yet. In this study, we systematically investigated the piezoelectric ($e_{11}$, $d_{11}$) and elastic (C$_{11}$ and C$_{12}$) properties of 128 thermodynamically stable two-dimensional (2D) semiconductor materials by employing first-principle methods. Our high-throughput approach demonstrates that the materials containing Group-\textrm{V} elements produce significantly high piezoelectric strain constants, $d_{11}$ $>$ 40 pmV$^{-1}$, and 49 of the materials considered have the $e_{11}$ coefficient higher than MoS$_{2}$ insomuch as BrSSb has one of the largest $d_{11}$ with a value of 373.0 pmV$^{-1}$.  Moreover, we established a simple empirical model in order to estimate the $d_{11}$ coefficients by utilizing the relative ionic motion in the unit cell and the polarizability of the individual elements in the compounds. 
\end{abstract}

\maketitle

\section{Introduction}

Piezoelectricity, defined as the electrical polarization of semiconductor materials without inversion symmetry in response to applied mechanical stress, is of utmost importance for sensing, actuating, and energy harvesting applications. Therefore, it has been attracted notable attention scientifically and technologically in the last decade. Recently, first-principles calculations on layered materials have triggered a great interest in piezoelectricity and its applications~\cite{Cui2018}. The piezoelectric strain coefficients ($d_{11}$), a measure of the mechanical to electrical energy conversion efficiency, of various two-dimensional materials such as single-layer nitrides, gapped graphene, transition metal dichalcogenides (TMDCs), transition metal dioxides (TMDOs), group \textrm{III} and \textrm{IV} monochalcogenides, and \textrm{II}-\textrm{VI} and \textrm{III}-\textrm{V}  compounds, have been determined as one or two orders of magnitude larger than that of traditionally used bulk materials such as $\alpha$-quartz ($d_{11}$ = 2.3~pmV$^{-1}$)~\cite{ref1,ref2}, $wurtzite$-GaN ($d_{33}$ = 3.1~pmV$^{-1}$)~\cite{ref3}, and $wurtzite$-AlN ($d_{33}$ = 5.1~pmV$^{-1}$)~\cite{ref3}. Considering various 2D materials, the values as high as 7.39, 8.47, 10.3, 16.3, 21.7, 27.3, 212.1, 250.5~pmV$^{-1}$ have been obtained for monolayers, MoTe$_2$~\cite{ref4}, CrSe$_2$~\cite{ref5}, MoSTe~\cite{ref6}, CrTe$_2$~\cite{ref5}, CdO~\cite{ref4}, SnOSe~\cite{ref7}, GeSe~\cite{ref8}, and SnSe~\cite{ref8}  crystals, respectively. In addition to these theoretical calculations, experimental studies have also demonstrated the intriguing piezoelectric response of 2D materials. Li \textit{et al.} indirectly verified the existence of monolayer $h$-BN  piezoelectricity~\cite{ref9} and Ares \textit{et al.} measured the $e_{11}$ coefficient of monolayer $h$-BN as 2.91$\times$10$^{-10}$ Cm$^{-1}$~\cite{ref10}. Wu \textit{et al.} experimentally affirmed piezoelectricity in free-standing monolayer MoS$_2$, they found that MoS$_2$ exhibits a strong piezoelectric effect for an odd number of layers in which case the inversion symmetry is broken~\cite{ref11}. Most recently Dai \textit{et al.} demonstrated that the piezoelectric properties of the monolayer MoS$_2$ can be tailored by the addition of grain boundaries~\cite{ref12}. Zhu \textit{et al.} measured the $e_{11}$ coefficient of MoS$_2$ as 2.9$\times$10$^{-10}$~Cm$^{-1}$, confirming the previous theoretical calculations~\cite{ref13}. Lu \textit{et al.} demonstrated the piezoelectric response in monolayer MoSSe~\cite{ref14}. Also, Zelisko \textit{et al.} determined the $e_{11}$ for \textit{g}-C$_{3}$N$_{4}$ as 2.18$\times$10$^{-10}$~Cm$^{-1}$~\cite{ref15}.

Indeed, these results have clearly revealed that layered materials possess high piezoelectric as well as flexoelectric responses, that makes them very attractive candidates for the next generation nanoscale technological applications such as stretchable smart electronics, skins, switches and many types of sensors. In fact, when the further enhancement of this exciting property via nanoengineering is considered~\cite{ref16,ref17} the importance of this field of research becomes more prominent.

On the other hand, these developments offer vast opportunities for computational material science since a large number of materials can now be analyzed and their properties determined without experimentation. For this purpose simple models can be developed using vast amount of data generated to estimate promising materials and properties. Practically, the information generated using computational techniques may potentially pave the way to rapid design and development based on the 2D piezoelectric/flexoelectric materials.

Considering that fact, we performed a high throughput  search study to predict the promising 2D hexagonal and trigonal crystals with large piezoelectric responses that can be exploited for device applications.  We selected thermodynamically stable semiconductor materials without inversion symmetry from the Computational 2D Materials Database (C2DB)~\cite{ref18}. Our first-principles calculations and empirical model provided a fundamental understanding of the correlation between piezoelectric properties of 128 monolayers and basic features (atomic mass and polarizability) of the constituent elements. Besides, we found that the relative motion of atoms within the monolayer against the external stress determines the size of net polarization and thereby piezoelectric coefficients.  As a result, we identified the most promising materials with high piezoelectric responses.  

\section{COMPUTATIONAL DETAILS}
First principle calculations based on density functional theory (DFT) were performed as implemented in the Vienna \textit{Ab initio} Simulation Package (VASP) code~\cite{ref19, ref20, ref21}. Exchange-correlation effects were included using  generalized gradient approximation (GGA) within Perdew-Burke-Ernzerh formalism (PBE)~\cite{ref22}. Single-electron wave functions were expanded up to an energy-cutoff of 700~eV for the structural relaxations, elastic and piezoelectric constant calculations. Brillouin-zone integrations were performed using a $\Gamma$-centered regular 24$\times$24$\times$1 $k$-point mesh within the Monkhorst–Pack~\cite{ref23} scheme.  The convergence criteria for the electronic and ionic relaxations were set to 10$^{-6}$~eV and 10$^{-2}$~eV\AA$^{-1}$ respectively. To prevent artificial interlayer interactions, the vacuum space was taken at least 20 {\AA} for all structures considered. Piezoelectric coefficient tensor $e_{ijk}$ and elastic stiffness tensor C$_{ijkl}$ were calculated by using density functional perturbation theory (DFPT)~\cite{ref24}.

\subsection{Theoretical Background}
The piezoelectric phenomenon is an electromechanical coupling and occurs only in certain non-centrosymmetric semiconducting materials where an electric dipole moment develops upon the application of stress or strain. The coefficients describing piezoelectric effect, namely $e_{ijk}$ and $d_{ijk}$, are given as
\numparts
\begin{eqnarray}
e_{ijk}=\frac{\partial\mathbf{P_\textit{i}}}{\partial\varepsilon_{\textit{jk}}}=\frac{\partial\mathbf{\sigma_{\textit{jk}}}}{\partial\mathrm{E_{\textit{i}}}}\\
d_{\textit{ijk}}=\frac{\partial\mathbf{P_\textit{i}}}{\partial\sigma_{\textit{jk}}}=\frac{\partial\mathbf{\varepsilon_{\textit{jk}}}}{\partial\mathrm{E_{\textit{i}}}}\\
e_{\textit{ik}}=d_{\textit{ij}}\mathrm{C}_{\textit{jk}}\label{eq1}
\end{eqnarray}
\endnumparts
where ($j$,$k$) = {(1,1), (2,2), (3,3), (1,2), (2,3), (3,1)}, $i$ = {1, 2, 3}, \textbf{P}$_{i}$, E$_{i}$, $\varepsilon_{jk}$ and $\sigma_{jk}$ are piezoelectric polarization, macroscopic electric field, strain and stress, respectively. In the Voigt notation, $e_{ijk}$ and $d_{ijk}$ are reduced to $e_{il}$ and $d_{il}$, respectively, where $l$ $\in$ {1, 2, ..., 6}.
\begin{figure}[h!]
\centering
\includegraphics[width=0.8\textwidth]{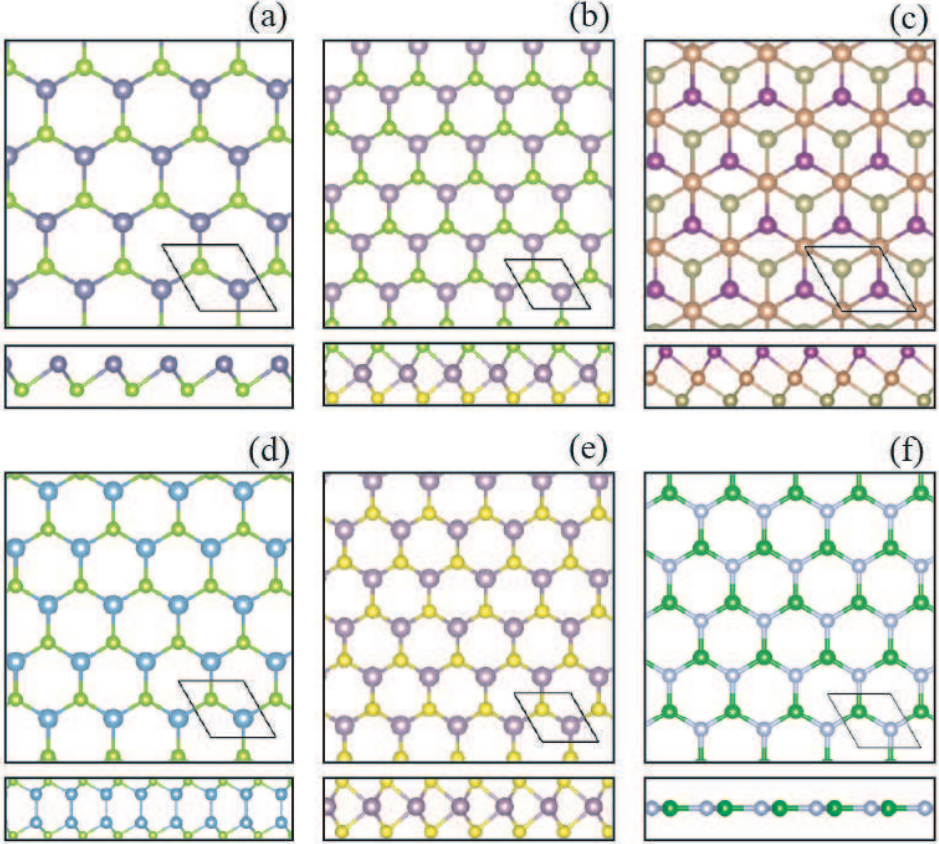}%
\caption{Top and side views of prototypes (a) PI, (b) PII, (c) PIII, (d) PIV, and (e) PV considered and the test case (f) $h$-BN.\label{fig1}}
\end{figure}

Based on the symmetry of the crystals, there exist certain amounts of independent elastic and piezoelectric constants. These independent constants depend on the point group symmetry of the crystals. Elastic response (elasticity), which is related to the mechanical properties of a material, can be defined as a reaction of the material on a macroscopic (or microscopic) scale to an external force~\cite{ref25}. Elasticity is an anisotropic property that is identified by 4$^{th}$ rank tensor and represented with a 6$\times$6 matrix~\cite{ref26}. On the other hand, the piezoelectric constant is a 3$^{rd}$ rank tensor and represented by a 3$\times$6 matrix in 3-dimensional space.

The 127 (plus $h$-BN) two-dimensional materials with hexagonal and trigonal symmetries are considered in this study. As listed in the tables depicted in \ref{App-I}, the 7, 43, 28, 13, and 36 of these materials can be classified into five different prototype structures, labeled as PI (named as GeSe prototype in C2DB), PII (named as MoSSe prototype in C2DB), PIII (named as BiTeI prototype in C2DB ), PIV (named as GaS prototype in C2DB ), and PV (named as MoS$_{2}$ prototype in C2DB), respectively. The crystal structures of these prototypes and test case $h$-BN are illustrated in Fig.~\ref{fig1} (a-f). While materials that belong to the PI-III have trigonal $3m$ symmetry, the materials that belong to the  PIV-V have hexagonal $\bar{6}m2$ symmetry. Hexagonal $\bar{6}m2$ exhibits mirror symmetry in the out of the plane direction, which nullifies the $e_{31}$, $d_{31}$, and C$_{14}$ constants. However, materials belong to PII-III prototypes include two different cation atoms with different atomic sizes and electronegativities, which gives rise to two inequivalent anion-cation bond lengths and charge distributions. As a result, the reflection symmetry is broken in the out-of-plane direction, resulting in a nonzero dipole moment and $e_{31}$ and $d_{31}$  are reflected in the piezoelectric tensors. But these constants have not been considered in this study due to the non-periodic nature of the two-dimensional materials investigated. The details of piezoelectric tensors corresponding to all crystal structure symmetries considered in this study are explained in \ref{App-I}.

Due to the considered vacuum layer of 3D periodic simulation cells in the $z$-direction, the coefficients were renormalized by multiplying with the $z$ lattice parameter to get 2D constants, \textit{i.e.}, C$_{ij}^{2D}$=$z$C$_{ij}^{3D}$  and $e_{ij}^{2D}$=$ze_{ij}^{3D}$ . This rescaling also changes the units of the constants from Nm$^{-2}$ to Nm$^{-1}$ and Cm$^{-2}$ to Cm$^{-1}$ for the C$_{ijk}$ and $e_{ik}$, respectively. Then the unit for $d_{ij}$ becomes mV$^{-1}$.

\section{RESULTS AND DISCUSSIONS}
\begin{figure}
\centering
\includegraphics[width=0.5\textwidth]{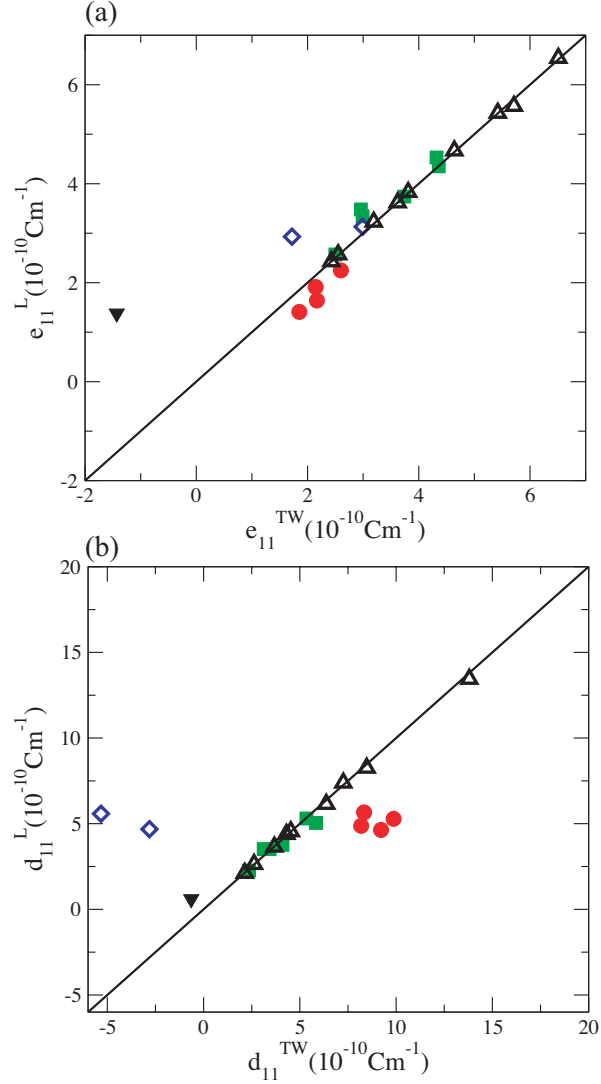}%
\caption{Comparison of the results calculated in this study and reported in the literature. The filled and empty symbols show the literature values obtained using BP and DFPT calculations, respectively. Superscripts L and TW represents literature and this work. MX$_{2}$ (where M= Cr, M, W and X= S,Se,Te),  MSSe (where M= Hf, Zr), MXY (where M= Mo,W and X,Y= S, Se, Te), MX (where M= Ge,Sn and X= S,Se) and $h$-BN represented by empty black triangle, empty blue diamond, solid green square, solid red dots and solid black triangle, respectively.\label{figure2}}
\end{figure}

As mentioned above, we investigated 128 (including $h$-BN) semiconductor two-dimensional materials without inversion symmetry, found as thermodynamically stable in the Computational 2D Materials Database (See also Tables S1-S5 in the Supplementary Information, SI). As expected, the piezoelectric properties of some of the materials considered in our calculations were previously investigated with density functional perturbation theory (DFPT) and/or Berry Phase (BP) approach~\cite{ref4,ref27,ref28,ref29,ref30}. The satisfactory agreement, depicted in Fig.~\ref{figure2} (See also Table S11 in SI), between our results and the ones published in the literature clearly demonstrate the accuracy of the approach used in this work.

Note that, for the materials presented by red dots (GeS, GeSe, SnS, and SnSe), there is a considerable difference between the calculated $d_{11}$ values, and the ones obtained by Hu and Dong~\cite{ref27}. In fact, the consistency in calculated $e_{11}$ values in these two cases demonstrates that the significant differences arise from the elastic constants, C$_{11}$ and C$_{12}$  in these two cases, yielding different $d_{11}$ values. Indeed, our elastic constant values are in good agreement with the values presented in C2DB database.

\subsection{Piezoelectric Properties}
When prototypes are compared, the materials belonging to the PII prototype yield larger piezoelectric constants as seen in \ref{App-II} (Table~\ref{Table-App-II}). The origin of this behavior is attributed to the relative motion of anions in the lattice against applied strain. For this prototype, both anions move in the same direction giving rise to increased polarization, and thereby a larger piezoelectric effect. In sharp contrast, the anions in monolayers with PIII structure move in the opposite directions, resulting in a reduced polarization, thereby smaller piezoelectric constants, see \ref{App-II} (Table~\ref{Table-App-I}). The amount of net polarization is sensitive to the atomic polarizability (P) to the mass (M) ratio of constituent atoms (polarizability to mass ratio P/M). This statement is discussed further in the following parts, but briefly, the individual atomic polarization to mass ratio can be used as a good indicator of the piezoelectric response.

The sub-group of materials with PII structure containing As and Sb cations yields the highest values, as seen in Table~\ref{table1}. Among them, the BrSSb possesses a $d_{11}$ value of 373.0 pmV$^{-1}$, which is extremely high when compared with the well known piezoelectric materials such as quartz. Within the materials that belong to PIII, the highest values are obtained for three materials comprising As and Sb cations as seen in Table~\ref{table1}. 
\begin{table}[h!]
\centering
\caption{The relaxed ion $e_{11}$ and $d_{11}$ coefficients for the most promising candidate monolayers.}
\label{table1}
\begin{tabular}{ |c|c|c| }
\hline
\hline
\multicolumn{3}{|c|}{\textbf{PII Prototype}} \\
\hline
\textbf{Material}& $e_{11}$(10$^{-10}$ Cm$^{-1}$) & $d_{11}$(pmV$^{-1}$) \\
\hline
BrSSb & 11.82 & 373.00 \\
AsBrTe & 59.66   & 298.44 \\
ClSbSe & 12.40 & 203.55 \\
\hline
\multicolumn{3}{|c|}{\textbf{PV Prototype}} \\
\hline
\textbf{Material}& $e_{11}$(10$^{-10}$ Cm$^{-1}$) & $d_{11}$(pmV$^{-1}$) \\
\hline
PbSe$_2$ & 2.60 & 114.63 \\
PbS$_2$ & 2.32   & 104.51 \\
SnI$_2$ & 4.93 & 102.75 \\
\hline
\multicolumn{3}{|c|}{\textbf{PIII Prototype}} \\
\hline
\textbf{Material}& $e_{11}$(10$^{-10}$ Cm$^{-1}$) & $d_{11}$(pmV$^{-1}$) \\
\hline
AsIS & 21.68 & 98.79 \\
AsBrS & 14.02   & 49.87 \\
BrSSb & 8.99   & 42.84 \\
\hline
\hline
\end{tabular}
\end{table}
The only difference in the chemical composition of these materials is the anions. Therefore, the difference in piezoelectric coefficients can be explained by the atomic polarization and mass ratio (P/M) values of I (0.259), Br (0.263), and Cl (0.422). One should note that the AsIS material exhibits the largest piezoelectric coefficient among these three materials, but the element I has the lowest P/M value. This is because the anions in materials belonging PIII move in the opposite directions. PbSe$_{2}$ structure has the exceptionally high $d_{11}$ value of 114.6 pmV$^{-1}$ within the materials belong to PV in which the average $d_{11}$ value is 12.8 pmV$^{-1}$ when excluding PbSe$_{2}$. However, the $e_{11}$ value of PbSe$_{2}$ is not that high when compared with the other materials in this prototype. The distinctly smaller C$_{11}$ to C$_{12}$ ratio (this is also the case for BrSSb structure in PII) results in an enhanced $d_{11}$ value for this monolayer. The materials classified as PIV give the lowest results with the average $d_{11}$ value of 2.2 pmV$^{-1}$ as seen in \ref{App-II} (Table~\ref{Table-App-IV}). Consequently, the systematic investigation  in this study clearly shows that the considered 19 materials possessing a giant piezoelectric effect (AsIS and ISSb with the PIII structure, BrSSb, AsBrTe, ClSbSe, ISSb, BrSbSe, AsBrSe, ISbSe, AsISe, BrSbTe, BiClS, ISbTe and BiIS with the PII structure, and PbSe$_{2}$, PbS$_{2}$, SnI$_{2}$, PbCl$_{2}$ and PbBr$_{2}$  with the PV structure). Also, 34, 47, and 72 different two-dimensional crystals with a piezoelectric strain coefficient greater than 20, 10, and 3.68 (that of MoS$_{2}$) pmV$^{-1}$ are determined, respectively. These results clearly show the potential of these materials worth investigating in experiments in regard to the broad range of applications such as nanosized sensors, piezotronics, and energy harvesting in portable electronic devices.

\subsection{Empirical Model}
In literature, there are some empirical models proposed to estimate the $d_{11}$ values of the materials, considering the ratio between the polarizability of anion to cation. In a previous work, we also applied the same approach to the Group II-VI two-dimensional honeycomb monolayers~\cite{sevik2016peculiar}. In general, these models fairly explain the trend. However, they do not consider the ionic motion due to the external effect, which has a notable influence on the final value, according to our analysis discussed below. Blonsky \textit{et al.}~\cite{ref4}, stated that for the MX$_{2}$ materials (where M = Cr, Mo, W, Nb, Ta, and X = S, Se, Te) there is a direct correlation between the polarizability ratio of anion and cation, \textit{i.e.}, P$_{anion}$/P$_{cation}$ (where P is the polarizability of an atom), and piezoelectric constants of these materials. However, the correlation is valid only when \textit{a distinct correlation constant} is used for each M atom. 

In this respect, we developed a different approach which includes the effect of the polarizability of the atoms and the ionic motion as well. This approach leads to further understanding of why some of the materials possess enhanced piezoelectric properties. Considering the mechanism of the piezoelectricity, which involves polarization of the crystal due to an external effect, four mechanisms emerge; I. The electronic component of polarization, II. Ionic motion, III. Molecular orientation, and IV. Mobile charge carriers under external field~\cite{ref32}. For the case of piezoelectric response, the electronic components and ionic motion mechanisms are the main mechanisms. 

\begin{figure}
\centering
\includegraphics[width=0.8\textwidth]{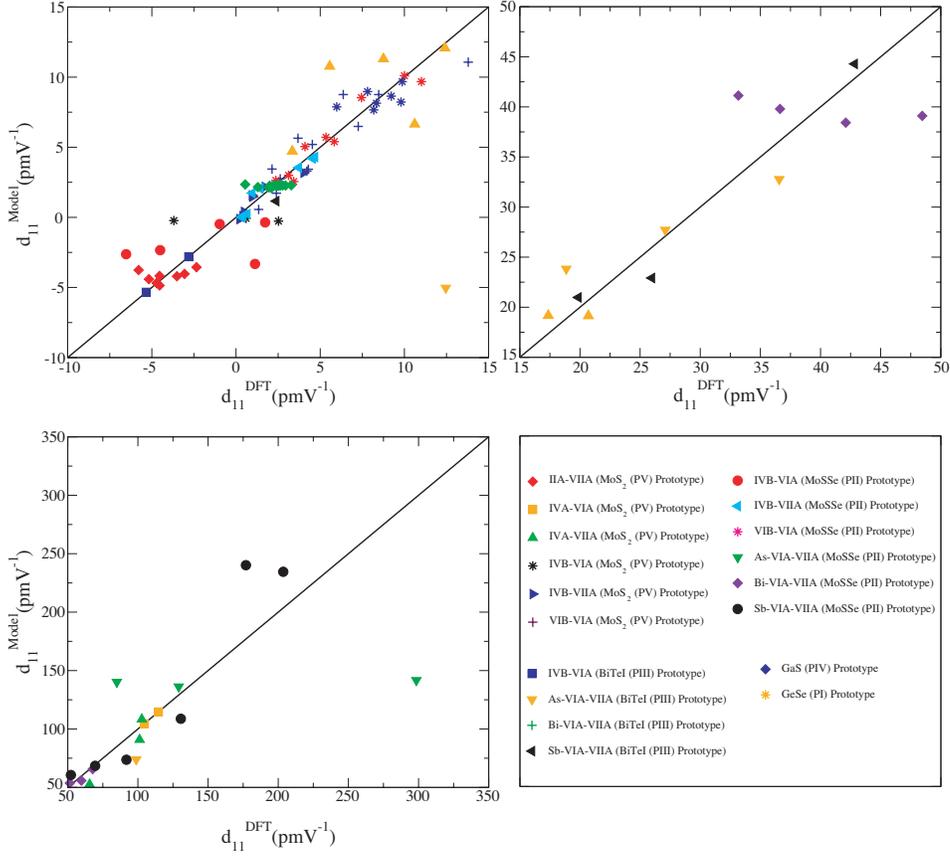}%
\caption{DFPT and Model comparison for $d_{11}$ of the materials.\label{figure3}}
\end{figure}

The electronic component of the polarization occurs when the electron cloud around the nucleus changes the orientation in favor of one way. This effect can be observed in all the constituent ions~\cite{ref32}. Therefore, we considered the atomic polarizability values in the proposed model. For the ionic motion mechanism, the situation is different. The material cannot experience a translation when subjected to deformation, meaning that the center of mass does not change when stress or strain is applied to the crystal, or under a piezoelectric response. Thus, the heavier atoms must move shorter relative to the lighter atoms to keep the center of mass unchanged. Since the ionic motion increases the polarization of a crystal, the heavier atoms contribute to overall polarization less due to their smaller ionic motion and therefore, the atomic mass is also considered. In addition, the bond strength also influences the ionic motion and thereby, we utilized the C$_{11}$ values of the materials as well.  To this end, considering three metrics, namely (1) polarizability~\cite{ref33} of the individual atoms (which has an increasing effect), (2) atomic mass, and (3) elastic constant C$_{11}$ of the crystal (which has a reducing impact on the $d_{11}$ values), we proposed a model as follows to estimate $d_{11}$ coefficients.
\begin{equation}
\label{eq18}
 d_{11} =m\frac{\textrm{P}}{\textrm{M}}\textrm{C}_{11}^{-1} + c.
\end{equation}
Where $m-c$, and C$_{11}$ are the correlation constants, and elastic constant, respectively. The $\frac{\textrm{P}}{\textrm{M}}$ corresponds to the "collective polarizability to mass ratio of the unit-cell", which is calculated by considering the number of anions and cations in the cell, and their relative motions induced by strain. Therefore, the $\frac{\textrm{P}}{\textrm{M}}$ ratios for different prototypes are calculated as  $\frac{\textrm{P}_{c}}{\textrm{M}_{c}} + \frac{\textrm{P}_{a_1}}{\textrm{M}_{a_1}} + \frac{\textrm{P}_{a_2}}{\textrm{M}_{a_2}}$ for PII and PV (contains 1 cation and 2 anions moving in the same direction), $\frac{\textrm{P}_{c}}{\textrm{M}_{c}} + \frac{\textrm{P}_{a}}{\textrm{M}_{a}}$ for PI (contains 1 cation and 1 anion), $2\frac{\textrm{P}_{c}}{\textrm{M}_{c}} + 2\frac{\textrm{P}_{a}}{\textrm{M}_{a}}$ for PIV (contains 2 cations and 2 anions moving in the same direction) and $\frac{\textrm{P}_{c}}{\textrm{M}_{c}} - \frac{\textrm{P}_{a_1}}{\textrm{M}_{a_1}} + \frac{\textrm{P}_{a_2}}{\textrm{M}_{a_2}}$ for PIII (contains 1 cation and 2 anions moving in the opposite direction). Fig.~\ref{figure3} presents the comparison of results predicted by our simple model and DFPT calculations. The strong correlation between the coefficients calculated with the proposed simple formula and DFPT is clear. Here, we used different correlation constants ($m$ and $c$) as listed in SI (Tables S6-S10 in SI) for different sub-groups due to the influence of other factors such as the orbital configurations of the constituent atoms and electronegativity.

\begin{figure}
\centering
\includegraphics[width=0.8\textwidth]{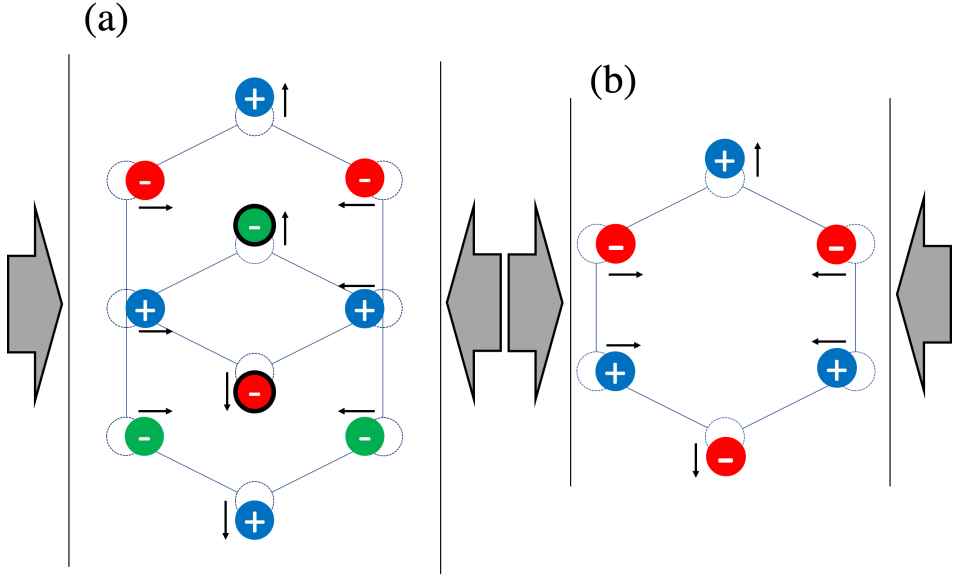}%
\caption{Schematic representation of the relative anion and cation pair motions; (a) for PIII, and (b) for all the other prototypes considered in this work. Here, the atoms represented with solid black circles correspond to the atoms moving in the opposite directions with respect to each other.\label{figure4}}
\end{figure}

However, the results depict that the piezoelectric coefficients of two-dimensional materials can be reasonably estimated with this simple formula which combines the atomic polarizability, atomic mass, and C$_{11}$. As seen in Fig.~\ref{figure3}, this model is valid for a very broad range of materials (see also Tables S6-S10 in SI). In order to change the polarization of the crystal, the cations and anions must move in the opposite directions as shown in Fig.~\ref{figure4}. However, in PIII, two anions in the unit-cell also move in the opposite directions, see the anions labeled with solid lines in Fig.~\ref{figure4} (a). This phenomenon hinders the piezoelectric response of the two-dimensional materials with the PIII structure. In this structure, the anions occupy the different lattice sites on the monolayer plane, and they move in the opposite direction against the external field (\textit{stress/strain}). Thus, we subtracted the contribution of the first anion since it moves in the opposite direction with respect to each other to the second one (the first anion is from \textrm{VII}A group element (Cl, I, Br) in the correlation equation).

\begin{table}[h!]
\centering
\caption{Comparison between BrSSb, ISSb and BrSbSe. P, p/m, $\frac{\textrm{P}}{\textrm{M}}$, $d_{11}^{p}$ and $d_{11}^{DFT}$ are the polarizability, polarizability to mass ratio of the ions, collective polarizability to mass ratio of the unit-cell (in equation \ref{eq18}), predicted $d_{11}$ and calculated $d_{11}$, respectively. For these materials, correlation constants m and c (in appropriate units) are 22943.92 and -722.98, respectively.}
\label{table2}
\begin{tabular}{ |c|c|c|c| }
\hline
\multicolumn{4}{|c|}{\textbf{BrSSb}} \\
\hline
 $\frac{\textrm{P}}{\textrm{M}}$ &  C$_{11}$ (Nm$^{-1}$) & $d_{11}^{DFT}$ (pmV$^{-1}$) & $d_{11}^{P}$ (pmV$^{-1}$) \\
\hline
1.221 & 27.08 & 373.00 & 311.45 \\
\hline
Element & Mass (g/mol)  &P (au) & p/m \\
\hline
Sb & 121.76 & 43.00  & 0.353 \\
Br & 79.90   & 21.00 & 0.263 \\
S & 32.07 & 19.40 & 0.605 \\
\hline
\hline
\multicolumn{4}{|c|}{\textbf{ISSb}} \\
\hline
 $\frac{\textrm{P}}{\textrm{M}}$ &  C$_{11}$ (Nm$^{-1}$) & $d_{11}^{DFT}$ (pmV$^{-1}$) & $d_{11}^{P}$ (pmV$^{-1}$) \\
\hline
1.217 & 29.00 & 176.96 & 240.12 \\
\hline
Element & Mass (g/mol) & P (au) &  p/m \\
\hline
Sb & 121.76 & 43.00  & 0.353 \\
I & 126.91   & 32.90 & 0.259 \\
S & 32.07 & 19.40 & 0.605 \\
\hline
\hline
\multicolumn{4}{|c|}{\textbf{BrSbSe}} \\
\hline
 $\frac{\textrm{P}}{\textrm{M}}$ &  C$_{11}$ (Nm$^{-1}$) & $d_{11}^{DFT}$ (pmV$^{-1}$) & $d_{11}^{P}$ (pmV$^{-1}$) \\
\hline
0.987 & 27.09 & 130.55 & 108.71 \\
\hline
Element& Mass\ (g/mol) & P (au) &  p/m \\
\hline
Sb & 121.76 & 43.00  & 0.353 \\
Br & 79.90   & 21.00 & 0.263 \\
Se & 78.96 & 28.90 & 0.366 \\
\hline
\hline
\end{tabular}
\end{table}

In general, using this simple approach, we are able to acquire a fundamental insight into why some materials have better piezoelectric responses. For instance, Table~\ref{table2} shows the materials, namely BrSSb, ISSb and BrSbSe, belonging to PII. One can see that the $\frac{\textrm{P}}{\textrm{M}}$ values are similar for the BrSSb and ISSb materials, but the large difference in the piezoelectric coefficient arises from the C$_{11}$ value, which is smaller for the BrSSb.

\section{CONCLUSION}
In this work, we report the elastic and piezoelectric properties of 128 thermodynamically stable 2D materials calculated. Our systematic analysis clearly shows that more than 70 crystals possess $\textit{d}_{11}$ coefficient values offering high potential for technological applications. Also, notable systems such as, AsIS and ISSb with the PIII structure, BrSSb, AsBrTe, ClSbSe, ISSb, BrSbSe, AsBrSe, ISbSe, AsISe, BrSbTe, BiClS, ISbTe and BiIS with the PII structure, and PbSe$_{2}$, PbS$_{2}$, SnI$_{2}$, PbCl$_{2}$ and PbBr$_{2}$  with the PV structure were determined to have coefficients comparable with the materials currently used in various scientific and technological fields. BrSSb is predicted to have a $\textit{d}_{11}$ coefficient of 373.0 pmV$^{-1}$ which is the one of the highest $\textit{d}_{11}$ coefficient reported so far. Here, we also developed a simple, yet effective model based on the relative ionic motions and atomic polarizabilities to predict the $\textit{d}_{11}$ coefficient of crystals considered. Our model provides an insight into which type of materials can have better piezoelectric coefficients. Furthermore, it is able to capture the nature of the predicted piezoelectric responses. The dynamic and thermodynamic stability of the considered materials are included in the 2D material database that shows a high potential of synthesizability of these types of materials. We believe that our findings can lead to new experimental studies to realize novel piezoelectric materials for various applications.

\ack
We acknowledge the support from the Eskisehir Technical University (ESTU-BAP 19ADP080). We also acknowledge financial support from ND EPSCoR through NSF grant OIA-1355466. Computational resources were provided by the High Performance and Grid Computing Center (TRGrid e-Infrastructure) of TUBITAK ULAKBIM, the National Center for High Performance Computing (UHeM) of Istanbul Technical University.

 \appendix
\section{The Details of Piezoelectric Tensor Calculations}\label{App-I}
The elastic and piezoelectric strain and stress tensors for trigonal $3m$, and hexagonal $\bar{6}m2$  symmetries can be described respectively as;
\begin{equation}
  \mathrm{C}_{jk} = \left(
  \begin{array}{*{6}{@{}C{0.8\mycolwd}@{}}}
\mathrm{C}_{11} & \mathrm{C}_{12} & \mathrm{C}_{13} & \mathrm{C}_{14} & \cdot & \cdot \\
\mathrm{C}_{12} & \mathrm{C}_{11} & \mathrm{C}_{13} & -\mathrm{C}_{14} & \cdot & \cdot \\
\mathrm{C}_{13} & \mathrm{C}_{13} & \mathrm{C}_{33} & \cdot & \cdot & \cdot \\
\mathrm{C}_{14} & -\mathrm{C}_{14} & \cdot & \mathrm{C}_{44} & \cdot & \cdot \\
\cdot & \cdot & \cdot & \cdot & \mathrm{C}_{44} & \mathrm{C}_{14} \\
\cdot & \cdot & \cdot & \cdot & \mathrm{C}_{14} & \frac{\mathrm{C}_{11}-\mathrm{C}_{12}}{2}
  \end{array}\right)
  \end{equation}
  \begin{equation}
    e_{ik} = \left(
  \begin{array}{*{6}{@{}C{0.8\mycolwdd}@{}}}
e_{11} & -e_{11} & \cdot & \cdot & e_{15} & \cdot \\
\cdot  & \cdot & \cdot & e_{15} & \cdot & -e_{11} \\
e_{31} & e_{31} &e_{33} & \cdot & \cdot  & \cdot 
  \end{array}\right)
  \end{equation}
  \begin{equation}
      d_{ij} = \left(
    \begin{array}{*{6}{@{}C{0.8\mycolwdd}@{}}}
d_{11} & -d_{11} & \cdot & \cdot & d_{15} & \cdot \\
\cdot  & \cdot & \cdot & e_{15} & \cdot & -2d_{11} \\
d_{31} & d_{31} &d_{33} & \cdot & \cdot  & \cdot 
  \end{array}\right)
  \end{equation}
and,
\begin{equation}
  \mathrm{C}_{jk} = \left(
  \begin{array}{*{6}{@{}C{0.8\mycolwd}@{}}}
\mathrm{C}_{11} & \mathrm{C}_{12} & \mathrm{C}_{13} & \cdot & \cdot & \cdot \\
\mathrm{C}_{12} & \mathrm{C}_{11} & \mathrm{C}_{13} & \cdot & \cdot & \cdot \\
\mathrm{C}_{13} & \mathrm{C}_{13} & \mathrm{C}_{33} & \cdot & \cdot & \cdot \\
\cdot & \cdot & \cdot & \mathrm{C}_{44} & \cdot & \cdot \\
\cdot & \cdot & \cdot & \cdot & \mathrm{C}_{44} & \cdot \\
\cdot & \cdot & \cdot & \cdot & \cdot & \frac{\mathrm{C}_{11}-\mathrm{C}_{12}}{2}
  \end{array}\right)
  \end{equation}
  \begin{equation}
    e_{ik} = \left(
  \begin{array}{*{6}{@{}C{0.8\mycolwdd}@{}}}
e_{11} & -e_{11} & \cdot & \cdot & \cdot & \cdot \\
\cdot  & \cdot & \cdot & \cdot & \cdot & -e_{11} \\
\cdot & \cdot & \cdot & \cdot & \cdot  & \cdot 
  \end{array}\right)
  \end{equation}
  \begin{equation}
    d_{ij} = \left(
  \begin{array}{*{6}{@{}C{0.8\mycolwdd}@{}}}
d_{11} & -d_{11} & \cdot & \cdot & \cdot & \cdot \\
\cdot  & \cdot & \cdot & \cdot & \cdot & -2d_{11} \\
\cdot & \cdot & \cdot & \cdot & \cdot  & \cdot 
  \end{array}\right)
  \end{equation}

The algorithms implemented in the VASP code calculate these tensors assuming periodic boundary conditions in 3D. But for 2D materials, stress and strain are constrained in the basal plane, nullifying the stress/strain components for $\sigma_3$/$\varepsilon_3$, $\sigma_4$/$\varepsilon_4$ and $\sigma_5$/$\varepsilon_5$. Therefore, one can obtain a 2D representation of the elastic constants and piezoelectric coefficients for matrices trigonal $3m$, and hexagonal $\bar{6}m2$  symmetries respectively as follows: 
\begin{equation}
  \mathrm{C}_{jk} = \left(
  \begin{array}{*{3}{@{}C{\mycolwd}@{}}}
\mathrm{C}_{11} & \mathrm{C}_{12} & \cdot \\
\mathrm{C}_{12} & \mathrm{C}_{11} & \cdot \\
\cdot & \cdot & \frac{\mathrm{C}_{11}-\mathrm{C}_{12}}{2}
  \end{array}\right)\label{eq2}
  \end{equation}
  \begin{equation}
    e_{ik} = \left(
  \begin{array}{*{6}{@{}C{\mycolwdd}@{}}}
e_{11} & -e_{11} & \cdot \\
\cdot  & \cdot & -e_{11} \\
e_{31} & e_{31} & \cdot 
  \end{array}\right)
  \end{equation}
  \begin{equation}
      d_{ij} = \left(
    \begin{array}{*{6}{@{}C{\mycolwdd}@{}}}
d_{11} & -d_{11} & \cdot  \\
\cdot  & \cdot & -2d_{11} \\
d_{31} & d_{31} & \cdot 
  \end{array}\right)
  \end{equation}
and,
\begin{equation}
  \mathrm{C}_{jk} = \left(
  \begin{array}{*{3}{@{}C{\mycolwd}@{}}}
\mathrm{C}_{11} & \mathrm{C}_{12} & \cdot \\
\mathrm{C}_{12} & \mathrm{C}_{11} & \cdot \\
\cdot & \cdot & \frac{\mathrm{C}_{11}-\mathrm{C}_{12}}{2}
  \end{array}\right)
  \end{equation}
  \begin{equation}
    e_{ik} = \left(
  \begin{array}{*{6}{@{}C{\mycolwdd}@{}}}
e_{11} & -e_{11} & \cdot \\
\cdot  & \cdot & -e_{11} \\
\cdot & \cdot & \cdot 
  \end{array}\right)
  \end{equation}
  \begin{equation}
      d_{ij} = \left(
    \begin{array}{*{6}{@{}C{\mycolwdd}@{}}}
d_{11} & -d_{11} & \cdot  \\
\cdot  & \cdot & -2d_{11} \\
\cdot  & \cdot  & \cdot 
  \end{array}\right)\label{eq3}
  \end{equation}
where, $i$, $j$, $k$ = { 1, 2, 3 }. Then from Eq.~\ref{eq3}, $d_{11}$ and $d_{31}$ become; 
\begin{eqnarray}
d_{11} &=& \frac{e_{11}}{\mathrm{C}_{11}-\mathrm{C}_{12}}\\
d_{31} &=& \frac{e_{31}}{\mathrm{C}_{11}+\mathrm{C}_{12}}
\end{eqnarray}
$d_{31}$ do not vanishes for monolayers with broken out-of-plane inversion symmetry. From Equations~\ref{eq2}-\ref{eq3}, one can see that only independent components are $e_{11}$, $d_{11}$, $e_{31}$, $d_{31}$, and C$_{11}$, C$_{12}$ for the structures considered in this study.

\newpage

\section{The Calculated Piezoelectric Stress and Strain Coefficients}\label{App-II}

\begin{table}[h!]
\centering
\begin{scriptsize}
\caption{\label{Table-App-I}The clamped and relaxed ion $e_{11}$ and $d_{11}$ coefficients of  the materials described with PIII.}
\begin{tabular}{ |c|c|c| }
\hline
\hline
\textbf{Material}& $e_{11}$(10$^{-10}$ Cm$^{-1}$) & $d_{11}$(pmV$^{-1}$) \\
\hline
AsBrS	&	14.02	&	49.87	\\
AsBrSe	&	6.19	&	18.86	\\
AsBrTe	&	1.02	&	3.15	\\
AsClS	&	11.77	&	36.57	\\
AsClSe	&	4.77	&	12.46	\\
AsClTe	&	-1.00	&	-3.01	\\
AsIS	&	21.68	&	98.79	\\
AsISe	&	8.54	&	27.09	\\
AsITe	&	2.79	&	9.04	\\
BrSbSe	&	5.10	&	19.87	\\
BrSbTe	&	1.56	&	6.41	\\
BrSSb	&	8.99	&	42.84	\\
ClSbSe	&	4.53	&	16.28	\\
ClSbTe	&	0.61	&	2.39	\\
ISbSe	&	6.19	&	25.98	\\
ISbTe	&	2.69	&	11.47	\\
ISSb	&	10.42	&	51.34	\\
BiBrS	&	4.45	&	17.36	\\
BiBrSe	&	3.06	&	12.41	\\
BiBrTe	&	1.23	&	5.56	\\
BiClS	&	4.14	&	15.72	\\
BiClSe	&	2.71	&	10.61	\\
BiClTe	&	0.78	&	3.35	\\
BiIS	&	5.13	&	20.70	\\
BiISe	&	3.67	&	15.54	\\
BiITe	&	1.87	&	8.75	\\
HfSSe	&	-1.72	&	-2.81	\\
ZrSSe	&	-2.99	&	-5.34	\\
\hline
\hline
\end{tabular}
\end{scriptsize}
\end{table}

\begin{table}
\centering
\begin{scriptsize}
\caption{\label{Table-App-II}The clamped and relaxed ion $e_{11}$ and $d_{11}$ coefficients of  the materials described with PII.}
\begin{tabular}{ |c|c|c| }
\hline
\hline
\textbf{Material}& $e_{11}$(10$^{-10}$ Cm$^{-1}$) & $d_{11}$(pmV$^{-1}$) \\
\hline
\hline
AsBrSe	&	16.56	&	129.05	\\
AsBrTe	&	59.66	&	298.44	\\
AsISe	&	13.43	&	84.97	\\
AsITe	&	9.90	&	48.70	\\
BrSbSe	&	9.82	&	130.55	\\
BrSbTe	&	8.25	&	69.54	\\
BrSSb	&	11.82	&	373.00	\\
ClSbSe	&	12.40	&	203.55	\\
ISbSe	&	8.98	&	91.77	\\
ISbTe	&	6.62	&	52.21	\\
ISSb	&	11.54	&	176.96	\\
BiBrS	&	6.70	&	59.73	\\
BiBrSe	&	5.98	&	48.46	\\
BiBrTe	&	5.03	&	36.63	\\
BiClS	&	7.10	&	67.78	\\
BiClSe	&	6.46	&	53.71	\\
BiClTe	&	5.90	&	40.65	\\
BiIS	&	6.71	&	51.47	\\
BiISe	&	5.75	&	42.10	\\
BiITe	&	4.54	&	33.17	\\
HfSeTe	&	-0.42	&	-0.97	\\
HfSSe	&	0.95	&	1.73	\\
TiSSe	&	0.63	&	1.13	\\
ZrSeTe	&	-1.76	&	-4.51	\\
ZrSTe	&	-2.56	&	-6.52	\\
HfBrCl	&	0.26	&	0.35	\\
HfBrI	&	0.27	&	0.44	\\
HfClI	&	0.45	&	0.67	\\
TiBrCl	&	2.33	&	3.73	\\
TiBrI	&	2.48	&	4.69	\\
TiClI	&	2.60	&	4.54	\\
ZrBrCl	&	0.62	&	0.97	\\
ZrBrI	&	0.79	&	1.43	\\
ZrClI	&	0.87	&	1.45	\\
CrSeTe	&	6.24	&	11.01	\\
CrSSe	&	5.64	&	7.45	\\
CrSTe	&	6.35	&	10.00	\\
MoSeTe	&	4.32	&	5.83	\\
MoSSe	&	3.74	&	4.10	\\
MoSTe	&	4.36	&	5.34	\\
WSeTe	&	2.96	&	3.43	\\
WSSe	&	2.50	&	2.37	\\
WSTe	&	2.99	&	3.12	\\
\hline
\hline
\end{tabular}
\end{scriptsize}
\end{table}

\begin{table}[h!]
\centering
\begin{scriptsize}
\caption{\label{Table-App-III}The clamped and relaxed ion $e_{11}$ and $d_{11}$ coefficients of  the materials described with PV.}
\begin{tabular}{ |c|c|c| }
\hline
\hline
\textbf{Material}& $e_{11}$(10$^{-10}$ Cm$^{-1}$) & $d_{11}$(pmV$^{-1}$) \\
\hline
BaBr$_2$	&	-0.44	&	-5.18	\\
BaI$_2$	&	-0.41	&	-5.79	\\
CaBr$_2$	&	-0.54	&	-3.06	\\
CaCl$_2$	&	-0.67	&	-3.51	\\
CaI$_2$	&	-0.37	&	-2.35	\\
SrBr$_2$	&	-0.57	&	-4.72	\\
SrCl$_2$	&	-0.65	&	-4.54	\\
SrI$_2$	&	-0.47	&	-4.55	\\
CrO$_2$ &	5.97	&	4.21	\\
CrS$_2$	&	5.42	&	6.37	\\
CrSe$_2$	&	5.71	&	8.47	\\
CrTe$_2$	&	6.51	&	13.79	\\
MoO$_2$	&	3.58	&	2.40	\\
MoS$_2$	&	3.62	&	3.68	\\
MoSe$_2$	&	3.81	&	4.53	\\
MoTe$_2$	&	4.64	&	7.26	\\
WO$_2$	&	2.36	&	1.35	\\
WS$_2$	&	2.43	&	2.14	\\
WSe$_2$	&	2.55	&	2.62	\\
WTe$_2$	&	3.19	&	4.30	\\
HfBr$_2$	&	0.15	&	0.22	\\
HfCl$_2$	&	0.32	&	0.40	\\
HfI$_2$	&	0.26	&	0.48	\\
TiBr$_2$	&	2.33	&	4.02	\\
ZrBr$_2$	&	0.58	&	0.97	\\
ZrCl$_2$	&	0.65	&	0.95	\\
ZrI$_2$	&	0.88	&	1.72	\\
HfTe$_2$	&	0.24	&	0.60	\\
TiSe$_2$	&	1.29	&	2.52	\\
ZrTe$_2$	&	-1.32	&	-3.69	\\
PbS$_2$	&	2.32	&	104.51	\\
PbSe$_2$	&	2.60	&	114.63	\\
PbCl$_2$	&	3.63	&	101.20	\\
PbBr$_2$	&	3.22	&	65.56	\\
PbI$_2$	&	2.82	&	39.78	\\
SnI$_2$	&	4.93	&	102.75	\\
\hline
\hline
\end{tabular}
\end{scriptsize}
\end{table}

\begin{table}[h!]
\centering
\begin{scriptsize}
\caption{\label{Table-App-IV}The clamped and relaxed ion $e_{11}$ and $d_{11}$ coefficients of  the materials described with PIV.}
\begin{tabular}{ |c|c|c| }
\hline
\hline
\textbf{Material}& $e_{11}$(10$^{-10}$ Cm$^{-1}$) & $d_{11}$(pmV$^{-1}$) \\
\hline
Al$_2$S$_2$	&	0.81	&	1.30	\\
Al$_2$Se$_2$	&	1.03	&	1.96	\\
Al$_2$Te$_2$	&	0.90	&	2.10	\\
Ga$_2$O$_2$	&	0.45	&	0.55	\\
Ga$_2$S$_2$	&	1.66	&	2.77	\\
Ga$_2$Se$_2$	&	1.70	&	3.28	\\
Ga$_2$Te$_2$	&	1.27	&	2.94	\\
In$_2$S$_2$	&	0.80	&	2.00	\\
In$_2$Se2$_2$	&	0.88	&	2.48	\\
In$_2$Te$_2$	&	0.78	&	2.53	\\
Tl$_2$S$_2$	&	0.69	&	2.38	\\
Tl$_2$Se$_2$	&	0.68	&	2.64	\\
Tl$_2$Te$_2$	&	0.50	&	2.23	\\
\hline
\hline
\end{tabular}
\end{scriptsize}
\end{table}

\begin{table}[h!]
\centering
\begin{scriptsize}
\caption{\label{Table-App-V}The clamped and relaxed ion $e_{11}$ and $d_{11}$ coefficients of  the materials described with PI.}
\begin{tabular}{ |c|c|c| }
\hline
\hline
\textbf{Material}& $e_{11}$(10$^{-10}$ Cm$^{-1}$) & $d_{11}$(pmV$^{-1}$) \\
\hline
GeS	&	2.60	&	8.34	\\
GeSe	&	2.15	&	8.18	\\
GeTe	&	1.47	&	5.98	\\
PbTe	&	1.39	&	9.80	\\
SnS	&	2.17	&	9.87	\\
SnSe	&	1.85	&	9.22	\\
SnTe	&	1.41	&	7.81	\\
\hline
\hline
\end{tabular}
\end{scriptsize}
\end{table}


\clearpage

\section*{References}

\begin{thebibliography}{35}%
\bibitem{Cui2018} 
Cui C, Xue F, Hu W J and Li L J 2018 Two-dimensional materials with piezoelectric and ferroelectric functionalities npj 2D Mater. Appl. 2 18

\bibitem{ref1} Bottom V E 1970 Measurement of the piezoelectric coefficient of quartz using the Fabry-Perot dilatometer J. Appl. Phys. 41 3941–4

\bibitem{ref2} Bechmann R 1958 Elastic and piezoelectric constants of alpha-quartz Phys. Rev. 110 1060–1

\bibitem{ref3} Lueng C M, Chan H L W, Surya C and Choy C L 2000 Piezoelectric coefficient of aluminum nitride and gallium nitride J. Appl. Phys. 88 5360–3

\bibitem{ref4} Blonsky M N, Zhuang H L, Singh A K and Hennig R G 2015 Ab Initio Prediction of Piezoelectricity in Two-Dimensional Materials ACS Nano 9 9885–91

\bibitem{ref5} Alyörük M M, Aierken Y, Çaklr D, Peeters F M and Sevik C 2015 Promising Piezoelectric Performance of Single Layer Transition-Metal Dichalcogenides and Dioxides J. Phys. Chem. C 119 23231–7

\bibitem{ref6} Yagmurcukardes M, Sevik C and Peeters F M 2019 Electronic, vibrational, elastic, and piezoelectric properties of monolayer Janus MoSTe phases: A first-principles study Phys. Rev. B 100 045415

\bibitem{ref7} Zhang X, Cui Y, Sun L, Li M, Du J and Huang Y 2019 Stabilities, and electronic and piezoelectric properties of two-dimensional tin dichalcogenide derived Janus monolayers J. Mater. Chem. C 7 13203–10

\bibitem{ref8} Fei R, Li W, Li J and Yang L 2015 Giant piezoelectricity of monolayer group IV monochalcogenides: SnSe, SnS, GeSe, and GeS Appl. Phys. Lett. 107 173104

\bibitem{ref9} Li Y, Rao Y, Mak K F, You Y, Wang S, Dean C R and Heinz T F 2013 Probing symmetry properties of few-layer MoS2 and h-BN by optical second-harmonic generation Nano Lett. 13 3329–33

\bibitem{ref10} Ares P, Cea T, Holwill M, Wang Y B, Roldán R, Guinea F, Andreeva D V., Fumagalli L, Novoselov K S and Woods C R 2020 Piezoelectricity in Monolayer Hexagonal Boron Nitride Adv. Mater. 32 1905504

\bibitem{ref11} Wu W, Wang L, Li Y, Zhang F, Lin L, Niu S, Chenet D, Zhang X, Hao Y, Heinz T F, Hone J and Wang Z L 2014 Piezoelectricity of single-atomic-layer MoS2 for energy conversion and piezotronics Nature 514 470–4

\bibitem{ref12} Dai M, Zheng W, Zhang X, Wang S, Lin J, Li K, Hu Y, Sun E, Zhang J, Qiu Y, Fu Y, Cao W and Hu P A 2020 Enhanced Piezoelectric Effect Derived from Grain Boundary in MoS2 Monolayers Nano Lett. 20 201–7

\bibitem{ref13} Zhu H, Wang Y, Xiao J, Liu M, Xiong S, Wong Z J, Ye Z, Ye Y, Yin X and Zhang X 2015 Observation of piezoelectricity in free-standing monolayer MoS2 Nat. Nanotechnol. 10 151–5

\bibitem{ref14} Lu A Y, Zhu H, Xiao J, Chuu C P, Han Y, Chiu M H, Cheng C C, Yang C W, Wei K H, Yang Y, Wang Y, Sokaras D, Nordlund D, Yang P, Muller D A, Chou M Y, Zhang X and Li L J 2017 Janus monolayers of transition metal dichalcogenides Nat. Nanotechnol. 12 744–9

\bibitem{ref15} Zelisko M, Hanlumyuang Y, Yang S, Liu Y, Lei C, Li J, Ajayan P M and Sharma P 2014 Anomalous piezoelectricity in two-dimensional graphene nitride nanosheets Nat. Commun. 5 1–7

\bibitem{ref16} Lu Y and Sinnott S B 2020 Density Functional Theory Study of Epitaxially Strained Monolayer Transition Metal Chalcogenides for Piezoelectricity Generation ACS Appl. Nano Mater. 3 384–90

\bibitem{ref17} Fan F R and Wu W 2019 Emerging Devices Based on Two-Dimensional Monolayer Materials for Energy Harvesting Research 2019 1–16

\bibitem{ref18} Haastrup S, Strange M, Pandey M, Deilmann T, Schmidt P S, Hinsche N F, Gjerding M N, Torelli D, Larsen P M, Riis-Jensen A C, Gath J, Jacobsen K W, Mortensen J J, Olsen T and Thygesen K S 2018 The Computational 2D Materials Database: High-throughput modeling and discovery of atomically thin crystals 2D Mater. 5

\bibitem{ref19} Kresse G and Furthmüller J 1996 Efficient iterative schemes for ab initio total-energy calculations using a plane-wave basis set Phys. Rev. B - Condens. Matter Mater. Phys. 54 11169–86

\bibitem{ref20} Blöchl P E 1994 Projector augmented-wave method Phys. Rev. B 50 17953–79

\bibitem{ref21} Joubert D 1999 From ultrasoft pseudopotentials to the projector augmented-wave method Phys. Rev. B - Condens. Matter Mater. Phys. 59 1758–75

\bibitem{ref22} Perdew J P, Burke K and Ernzerhof M 1996 Generalized gradient approximation made simple Phys. Rev. Lett. 77 3865–8

\bibitem{ref23} Kresse G and Joubert D 1999 From ultrasoft pseudopotentials to the projector augmented-wave method Phys. Rev. B - Condens. Matter Mater. Phys. 59 1758–75

\bibitem{ref24} Wu X, Vanderbilt D and Hamann D R 2005 Systematic treatment of displacements, strains, and electric fields in density-functional perturbation theory Phys. Rev. B - Condens. Matter Mater. Phys. 72 035105

\bibitem{ref25} Yu R, Zhu J and Ye H Q 2010 Calculations of single-crystal elastic constants made simple Comput. Phys. Commun. 181 671–5

\bibitem{ref26} Nye J F Physical Properties of Crystals: Their Representation by Tensors and Matrices

\bibitem{ref27} Hu T and Dong J 2016 Two new phases of monolayer group-IV monochalcogenides and their piezoelectric properties Phys. Chem. Chem. Phys. 18 32514–20

\bibitem{ref28} Dong L, Lou J and Shenoy V B 2017 Large In-Plane and Vertical Piezoelectricity in Janus Transition Metal Dichalchogenides ACS Nano 11 8242–8

\bibitem{ref29} Dimple, Jena N, Rawat A, Ahammed R, Mohanta M K and De Sarkar A 2018 Emergence of high piezoelectricity along with robust electron mobility in Janus structures in semiconducting Group IVB dichalcogenide monolayers J. Mater. Chem. A 6 24885–98

\bibitem{ref30} Duerloo K A N, Ong M T and Reed E J 2012 Intrinsic piezoelectricity in two-dimensional materials J. Phys. Chem. Lett. 3 2871–6

\bibitem{ref31} Råsander M and Moram M A 2015 On the accuracy of commonly used density functional approximations in determining the elastic constants of insulators and semiconductors J. Chem. Phys. 143 144104

\bibitem{sevik2016peculiar} Sevik C, Çaklr D, Gülseren O and Peeters F M 2016 Peculiar Piezoelectric Properties of Soft Two-Dimensional Materials J. Phys. Chem. C 120 13948–53

\bibitem{ref32} Newnham R E Properties of Materials: Anisotropy, Symmetry, Structure

\bibitem{ref33} Schwerdtfeger P and Nagle J K 2019 2018 Table of static dipole polarizabilities of the neutral elements in the periodic table* Mol. Phys. 117 1200–25

\end{thebibliography}
\end{document}